\documentclass[doublecol,figures]{epl2} 
\usepackage{amssymb,amsmath}
\usepackage{url}

\usepackage{epsfig}
\usepackage{amsmath}
\usepackage{amsfonts}
\usepackage{graphicx}
\usepackage{units}

\newcommand{\ket}[1]{|{#1}\rangle}                    
\newcommand{\bra}[1]{\langle{#1}|}

\newcommand{\kk}{\mathbf{k}}

\newcommand{\KK}{\mathbf{K}}

\newcommand{\qq}{\mathbf{q}}
\newcommand{\OO}{\mathbf{0}}

\newcommand{\uu}{\hat{u}}
\newcommand{\dd}{\hat{d}}
\newcommand{\bb}{\hat{b}}
\newcommand{\kf}{k_{\rm F}}
\newcommand{\Ef}{E_{\rm F}}
\newcommand{\ka}{\kappa}
\newcommand{\krel}{{k_{\rm rel}}}

\newcommand{\be}{\begin{equation}}
\newcommand{\ee}{\end{equation}}
\newcommand{\bea}{\begin{eqnarray}}
\newcommand{\eea}{\end{eqnarray}}

\title{Energy, decay rate, and effective masses for a moving polaron \\in a Fermi sea: Explicit results in the weakly attractive limit}
\shorttitle{Moving polaron in a Fermi sea: Explicit results in the weakly attractive limit}
\author{Christian Trefzger and Yvan Castin}  
\shortauthor{Christian Trefzger and Yvan Castin}  
\institute{Laboratoire Kastler Brossel, \'Ecole Normale Sup\'erieure and CNRS, UPMC, 24 rue Lhomond, 75231 Paris, France}


\abstract{We study the properties of an impurity of mass $M$ moving through a spatially homogeneous three-dimensional fully polarized Fermi gas of particles of mass $m$. In the weakly attractive limit,
where the effective coupling constant $g\to0^-$ and perturbation theory can be used, both for a broad and a narrow Feshbach resonance, we obtain an explicit analytical expression for  the complex energy $\Delta E(\KK)$ 
of the moving impurity up to order two included in $g$. This also gives access to its longitudinal and transverse effective masses 
$m_\parallel^*(\KK)$, $m_\perp^*(\KK)$, as functions of the impurity wave vector $\KK$. 
Depending on the modulus of $\KK$ and on the impurity-to-fermion mass ratio $M/m$ we identify four regions separated by singularities
in derivatives with respect to $\KK$ of the second-order term of $\Delta E(\KK)$, and we discuss the physical origin of these regions. Remarkably, 
the second-order term of $m_\parallel^*(\KK)$ presents points of non-differentiability, replaced by a logarithmic divergence for $M=m$, when $\KK$ 
is on the Fermi surface of the fermions. We also discuss the third-order contribution and relevance for cold atom experiments.}

\begin{document}
\maketitle
\date{\today}

\section{Introduction}
Recent cold atom experiments have reached an unprecedented accuracy in measuring the equation of state of an 
 interacting Fermi gas~\cite{Navon,Nascimbene,Zwierlein,Kris}. This has allowed to confirm that in strongly spin-polarized configurations 
the minority atoms dressed by the Fermi sea of the majority atoms form a normal gas of
quasiparticles called Fermi polarons~\cite{Lobo,Chevy,Svistunov}. 

While the problem of a single impurity at rest in a Fermi sea has been thoroughly studied~\cite{Walecka, BishopNucl, Bishop,Lobo,Chevy,Svistunov,Combescot,MassignanBruun,TCPRA,VanHoucke,revue_Massignan}, previous works on a moving impurity
have focused only on its decay rate~\cite{Bishop,Stringari}. However, the real part of the polaronic complex energy is also of theoretical interest and is experimentally accessible~\cite{Zaccanti}.
Whereas most theories for the impurity at rest successfully used a variational ansatz~\cite{Chevy}, this ansatz is not
reliable anymore for a moving polaron with momentum $\hbar \KK$ since at 
low $K$ it wrongly predicts a zero decay rate in contrast to the $K^4$-law found in~\cite{Bishop,Stringari}. 
 
\begin{figure}[tbp]
\begin{center}
\includegraphics[width=0.9\linewidth,clip=]{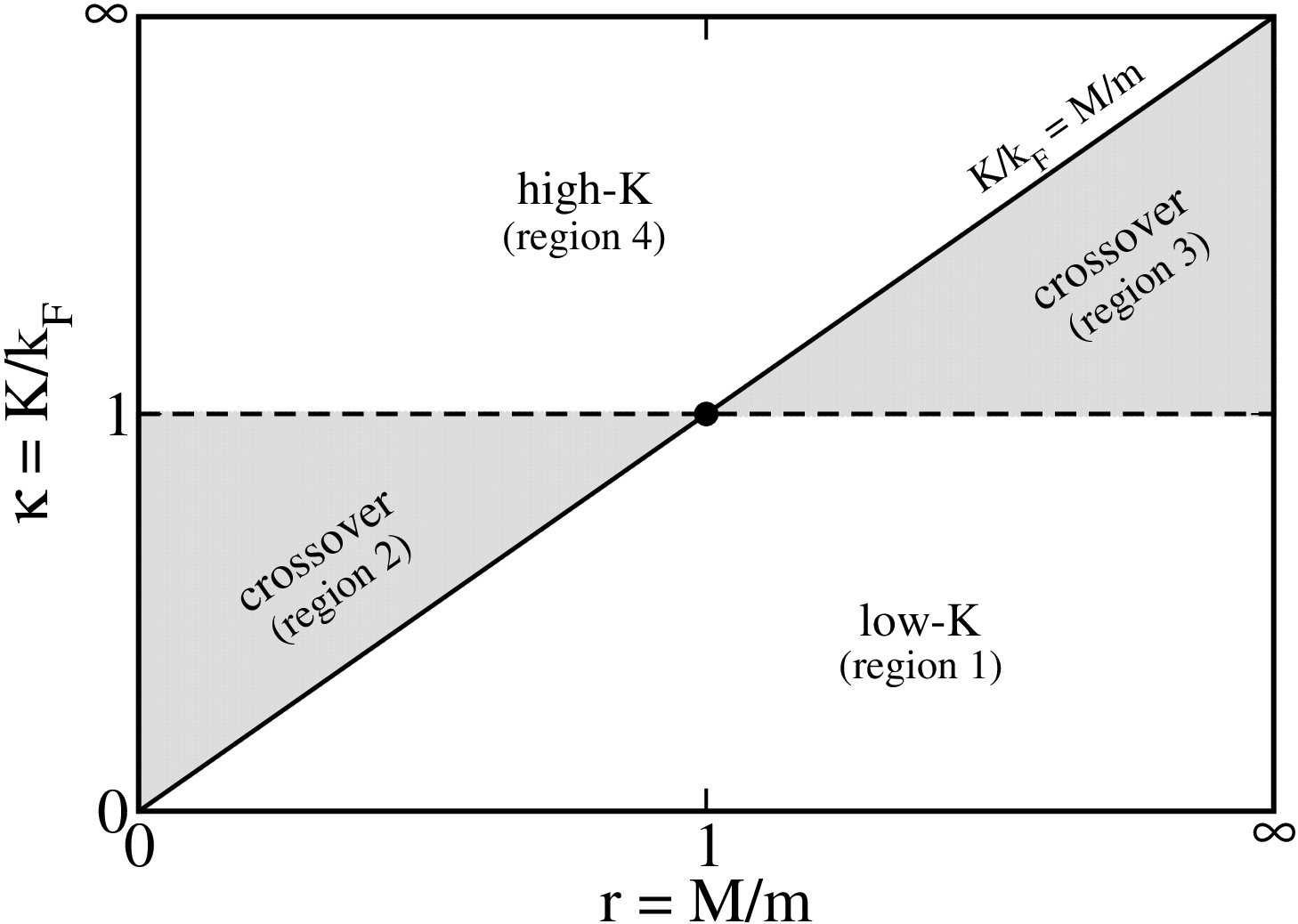}
\caption{Regions of the moving polaron in the  $M/m$ -- $K/\kf$ plane, where the second-order correction to the complex energy is a smooth
function of $K/\kf$. On the dashed line, $K=\kf$,
and on the solid line, $K=\kf M/m$, a logarithmic divergence (discontinuity) appears in the $n$th order derivative with respect to $K$ 
of the real part (imaginary part) of the second-order term of the polaronic complex energy~(\ref{eq:ener_pol}): $n=3$ for the dashed line, $n=4$ for the solid line,  $n=2$
at the point ($M=m$,$K=\kf$).} 
\label{fig:regions}
\end{center}
\end{figure}

In this work, we focus on the weakly attractive regime $\kf a\to0^-$, where $a$ is the $s$-wave scattering length of a minority 
atom and a fermion and $\kf$ is the Fermi wave number of the majority atoms.
Using a systematic expansion in powers of $\kf a$, 
we go beyond the Fermi liquid description of~\cite{Stringari}:  
We determine not only the decay rate, but also the real part of the polaronic complex energy, and being not restricted to
low momenta, we have access to momentum-dependent effective masses of the polaron. 
Within our microscopic approach we have a complete description of the system:
We determine regions of parameters separated by singularities in derivatives of the second-order term of the polaronic complex energy,
as shown in Fig.\ref{fig:regions}. 
We discuss the physical origin of the singularities and the relevance for current cold atom experiments.

\section{The model}
\label{sec:model}
At zero temperature, we consider in three dimensions an ideal Fermi gas of particles of same spin state and mass $m$ perturbed by the presence
of a moving impurity of mass $M$ and momentum $\hbar \KK$. While we assume no interactions among the fermions (contrary to~\cite{Nishida}), the impurity interacts with each 
fermion through a $s$-wave interaction of negligible range. 
The system is enclosed in a quantization volume $\mathcal{V}$ with periodic boundary conditions, and is described by the Hamiltonian $\hat{H} = \hat{H}_0 + \hat{V}$:
\begin{eqnarray}
\label{eq:H0}
\hat{H}_0 &=& \sum_\kk \left(\varepsilon_\kk \uu^\dag_\kk \uu_\kk + E_\kk \dd^\dag_\kk \dd_\kk \right), \\
\hat{V}     &=& \frac{g_0}{\mathcal{V}}\sum_{\substack{\kk_1,\kk_2 \\ \kk_3,\kk_4}} \delta^{\rm mod}_{\kk_1+\kk_2,\kk_3+\kk_4}
\dd^\dag_{\kk_4} \uu^\dag_{\kk_3} \uu_{\kk_2} \dd_{\kk_1}, 
\label{eq:V}
\end{eqnarray}
where $\hat{u}_\kk$ ($\hat{d}_\kk$) annihilates a fermion (impurity) with wave vector $\kk$ and kinetic energy $\varepsilon_\kk = \frac{\hbar^2 k^2}{2m}$
($E_\kk = \frac{\hbar^2 k^2}{2M}$). 
To avoid ultraviolet divergences we use a standard lattice model where the positions of the particles are discretized on a cubic lattice of elementary step $b\ll\kf^{-1}$ and binary interactions between the impurity and a fermion take place at the same lattice site with a bare coupling constant $g_0$. In momentum space this takes the form Eq.~(\ref{eq:V}). The model provides automatically a cutoff in the Fourier space since wave vectors are restricted to the first Brillouin zone. 
In Eq.~(\ref{eq:V}), the modified Kronecker delta is to be understood modulo a vector of the reciprocal
lattice and ensures conservation of the quasi-momentum. The bare coupling constant 
$g_0$ is then related to the scattering length $a$ through the formula\cite{Castin}
\be
\label{eq:bare}
\frac{1}{g_0} = \frac{1}{g} - \int_\mathrm{FBZ} \frac{d^3k}{(2\pi)^3} \frac{2\mu}{\hbar^2 k^2} \;\;\; \mbox{and} \;\;\; g=\frac{2\pi\hbar^2 a}{\mu},
\ee
where the integral is taken in the first Brillouin zone, $\mathrm{FBZ} = [-\pi/b, \pi/b[^3$, and $\mu=mM/(m+M)$ is the reduced mass.
In practice, at the end of the perturbative calculations to come, we will take the thermodynamic limit, where $\mathcal{V}\to+\infty$, and 
also the continuous space limit $b\to0^+$, so that the first Brillouin zone will converge to the whole Fourier space.

The goal is to calculate the complex energy of an impurity of wave vector $\KK$ moving in the zero temperature Fermi gas of $N$ particles. The ground state of
the $N$ fermions is the usual Fermi sea $\ket{\rm FS:N}$ of energy $E_{\rm FS}(N)$. In the absence of interactions the ground state
of the system is given by the free impurity immersed in the Fermi sea
\be
\label{eq:unperturbed_state}
\ket{\psi^{(0)}(\KK)} = \hat{d}_\KK^\dag \ket{\rm FS:N},
\ee
with an energy, measured with respect to the energy of the Fermi sea, given by
\be
\Delta E^{(0)}(\KK) = \bra{\psi^{(0)}(\KK)} \hat{H}_0 \ket{\psi^{(0)}(\KK)} - E_{\rm FS}(N) = E_\KK.
\ee 
We now consider the weakly interacting case $a\to0^-$, where a quasiparticle polaronic ground state is known to exist~\cite{Lobo,Chevy,Combescot,Svistunov}. For a fixed lattice spacing $b$, in the limit where $|a| \ll b$ we can Taylor expand the bare coupling constant (\ref{eq:bare}):
\be
\label{eq:expansion_g0}
g_0 \underset{g\to0^-}{=} g + g^2 \int_\mathrm{FBZ} \frac{d^3k}{(2\pi)^3} \frac{2\mu}{\hbar^2 k^2} + O\left(g^3\right),
\ee
and in this weakly attractive limit one can treat $\hat{V}$ in the Hamiltonian with usual perturbation theory.
Up to second order in $\hat{V}$, we find in the thermodynamic limit 
that the energy of the moving polaron is complex:
\be
\label{eq:ener_boite}
\Delta E(\KK) \underset{g\to0^-}{=} E_\KK + \rho g_0 -
\int_{\mathrm{FBZ}^2}'  \!\! \frac{d^3k d^3q}{(2\pi)^6} \frac{g_0^2}{F_{\kk,\qq}(\KK)} + O(g_0^3),
\ee
where the prime on the integral means 
that it is restricted to $q<\kf$ and to $k>\kf$, and we have introduced the mean density of the Fermi sea $\rho=N/\mathcal{V}$ related to the
Fermi wave number $\kf$ by the usual relation $\kf= (6\pi^2\rho)^{1/3}$, $\Ef=\frac{\hbar^2 \kf^2}{2m}$, and we have defined
\be
\label{eq:Fkq_def}
F_{\kk,\qq}(\KK) = E_{\KK+\qq-\kk}+\varepsilon_\kk-\varepsilon_\qq-E_\KK - i\eta,
\ee
with $\eta~\to~0^+$.
Substituting Eq.~(\ref{eq:expansion_g0}) into Eq.~(\ref{eq:ener_boite}) and keeping only 
terms up to order $g^2$,  as announced we get a finite expression for a vanishing lattice spacing
\be
\label{eq:ener_pol}
\Delta E(\KK) \underset{g\to0^-}{=} E_\KK + \rho g + \frac{(\rho g)^2}{\Ef} f(K/\kf) + O(g^3).
\ee
The dimensionless complex function $f\left(K/\kf\right)$ is an isotropic function of $\KK$ defined as follows:
\begin{multline}
\label{eq:integral}
\!\!\!\!\!  f\left(\frac{K}{\kf}\right) =\frac{\Ef}{\rho^2}\int'  \!\! \frac{d^3q}{(2\pi)^3} \int  \!\! \frac{d^3k}{(2\pi)^3} 
\left[  \frac{2\mu}{\hbar^2 k^2} - \frac{Y(k-\kf)}{F_{\kk,\qq}(\KK)}\right],
\end{multline}
where $Y(x)$ is the Heaviside function, and the integral
over $\kk$ is now taken over the whole Fourier space.

\section{Results}
The integral in Eq.~(\ref{eq:integral}) can be calculated exactly, as it will be detailed elsewhere~\cite{TCTC}. Using the Dirac relation 
$\lim_{\eta\to0^+} (x-i\eta)^{-1} = P\frac{1}{x} + i\pi \delta(x)$,
where $P$ is the principal value and $\delta(x)$ the Dirac distribution, we can separate the integral 
into its real and imaginary parts. 

The imaginary part of Eq.~(\ref{eq:ener_pol})
reveals that due to scattering of the impurity with the fermions, a moving polaron 
radiates particle-hole pairs~\cite{Stringari} and thus decays out of its initial momentum channel, at a rate $\Gamma_0$
given by
\be
\frac{\hbar \Gamma_0}{2} \underset{g\to0^-}{=} -\frac{(\rho g)^2}{\Ef} \Im f(K/\kf) + O(g^3).
\ee
\begin{figure}[t]
\begin{center}
\includegraphics[width=0.95\linewidth,clip=]{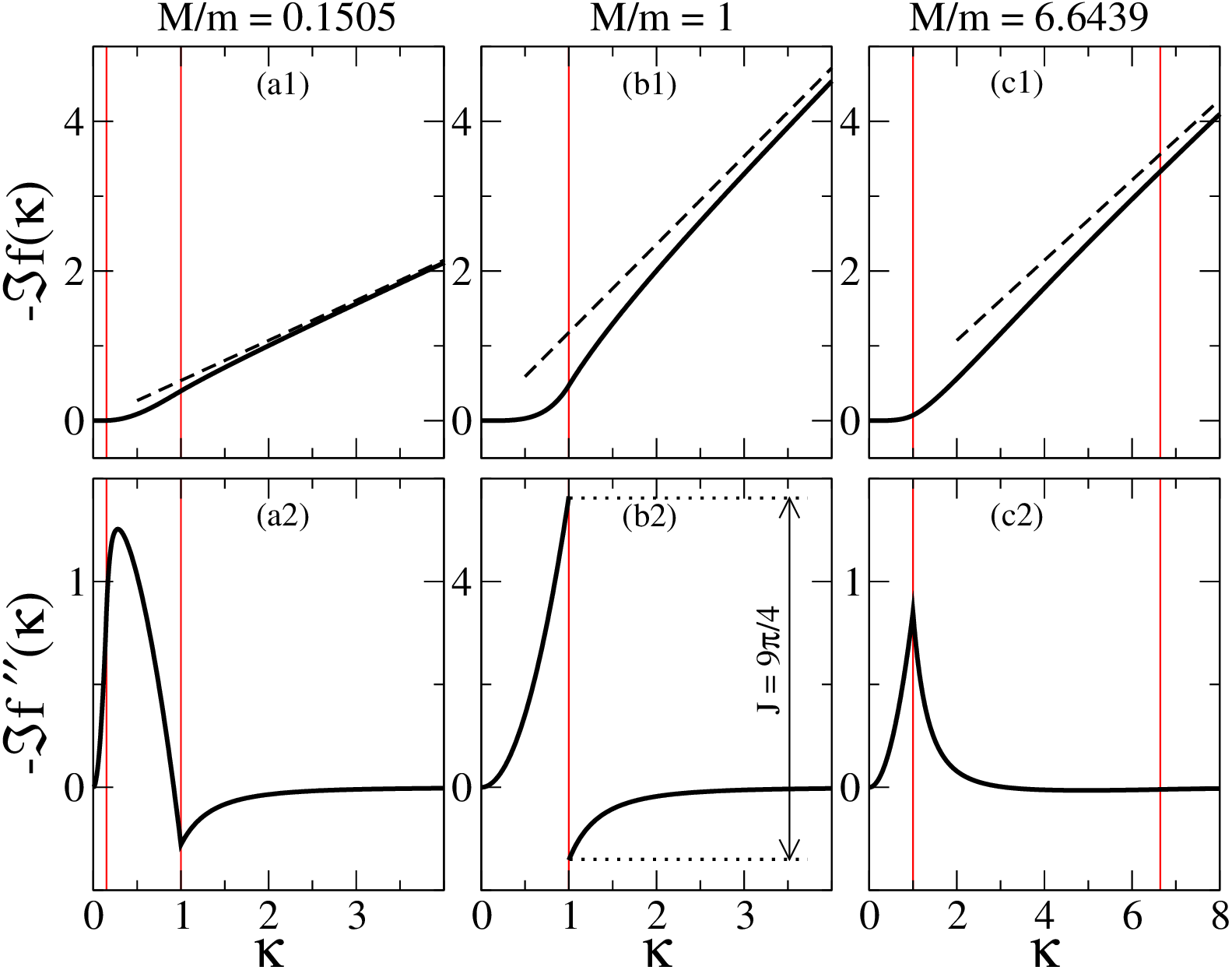}
\caption{(Color online) Function $-\Im f(\ka)$ entering the imaginary part of Eq.~(\ref{eq:ener_pol}) (row 1), and its second derivative 
entering the imaginary part of the inverse effective mass Eq.~(\ref{eq:invmeff}) (row 2), for various mass ratios. Dashed lines correspond to the 
asymptote of Eq.~(\ref{eq:Im3}). Red vertical lines correspond to
the boundaries at $\ka=1$ and at $\ka=M/m$.}
\label{fig:im-eff}
\end{center}
\end{figure}
This polaron decay rate coincides with the Fermi golden rule result, as best seen from Eq.~(\ref{eq:ener_boite}).
Depending on the impurity-to-fermion mass ratio $r=M/m$ and on the modulus of $\KK$, $\Im f(K/\kf)$  assumes different 
piecewise smooth (indefinitely derivable) functional forms on the four different regions represented in Fig.~\ref{fig:regions}. (i) The low-$K$ region, $0<K/\kf<\min(1,r)$, where
\be
\label{eq:region1}
-\Im f(\ka) \underset{\mathrm{region}\,1}{=} \frac{3\pi}{20 r}\ka^4.
\ee
As expected, $\Im f(0)= 0$. 
(ii) The $r<1$ (light impurity) crossover region, $\min(1,r)<K/\kf<\max(1,r)$, where
\begin{multline}
-\Im f(\ka) \underset{\mathrm{region}\,2}{=} \frac{3 \pi r}{20 (r^2-1)^2} \\ 
\times \left[ (r^2-2)\ka^4+10\ka^2-20r\ka+15r^2-4r^3/\ka \right].
\end{multline}
(iii) The $r>1$ (heavy impurity) crossover region, $\min(1,r)<K/\kf<\max(1,r)$, where
\begin{multline}
-\Im f(\ka) \underset{\mathrm{region}\,3}{=} \frac{3 \pi r}{20(r^2-1)^2}  \\ 
\times \left[\frac{\ka^4}{r^2} -10\ka^2+10(r^2+1)\ka-15r^2+\frac{6r^2-2}{\ka} \right].
\end{multline}
\begin{figure}[t]
\begin{center}
\includegraphics[width=0.95\linewidth,clip=]{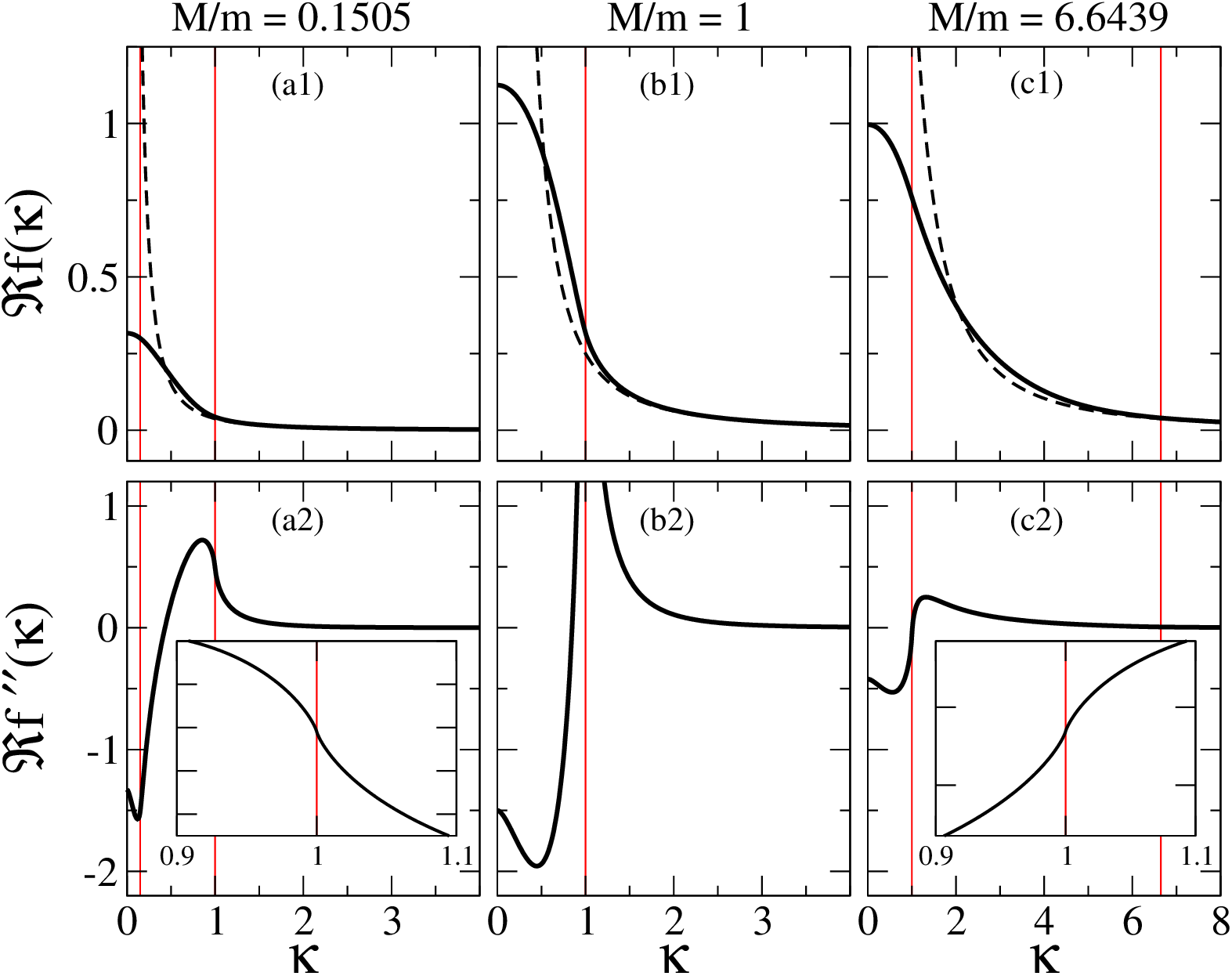}
\caption{(Color online) Function $\Re f(\ka)$ entering the real part of Eq.~(\ref{eq:ener_pol}) (row 1), and its second derivative 
entering the real part of the inverse effective mass Eq.~(\ref{eq:invmeff}) (row 2), for various mass ratios. Dashed lines:   
 Asymptotic behavior~(\ref{eq:re-large}). Red vertical lines: Boundaries at $\ka=1$ and at $\ka=M/m$. Insets: Enlargements showing the non-differentiability of the second-order term of $M/m_\parallel^*(\KK)$ at $K=\kf$.}
\label{fig:re-eff}
\end{center}
\end{figure}
There is no crossover region for a unit impurity-to-fermion mass ratio, $r=1$. 
(iv) Finally the high-$K$ region, $\max(1,r)<K/\kf$, where
\be
\label{eq:Im3}
- \Im f(\ka) \underset{\mathrm{region}\,4}{=} \frac{3 \pi r}{10 (r+1)^2} \frac{5\ka^2-2r-1}{\ka}.
\ee
Eqs.~(\ref{eq:region1},\ref{eq:Im3}) generalize to $r\neq1$ the results of~\cite{Bishop}. 
{Eq.~(\ref{eq:region1}) differs from the $K\to 0$ Fermi liquid theory of \cite{Stringari}
by a numerical coefficient, but \cite{Stringari} considers momentum rather than population decay.}
In Fig.~\ref{fig:im-eff} we plot the function $-\Im f(\ka)$ for different values of the mass ratio between the impurity 
and the fermions. 

For a large impurity wave vector, $K/\kf \gg \max(1,r)$, the asymptotic value of $- \Im f(\ka)$ is given by the asymptotic expansion
of Eq.~(\ref{eq:Im3}), and is linear in $K$.
This can be quantitatively recovered with a simple kinetic argument: For $K\gg(r+1)\kf$, due to conservation of energy and momentum,
the fermionic wave number $k$ after scattering with the impurity is $\gg\kf$ except in a small cone in the forward direction, so that the Pauli blocking
effect may be neglected. Also the impurity-to-fermion relative velocity is $v\simeq\hbar K/M$, which leads to a collision rate
$\Gamma_0 \simeq \rho \sigma v$, where $\rho$ is the fermionic density and $\sigma=4\pi a^2$ is impurity-to-fermion scattering cross section.

Let us now discuss the real part of the polaronic complex energy~(\ref{eq:ener_pol}) for which much less was known. 
We find that
\begin{multline}
\label{eq:Real_f}
\Re f(\ka) = \frac{3r}{r+1} - \frac{3r}{20(r^2-1)^2}\bigg\{ {\bigg[}\frac{(r^2-1)}{2}(\ka^2+20r-9) \\
+ \frac{(\ka-1)^3}{\ka}[\ka(\ka+3)(r^2-2)+6r^2-2]\ln \left| \frac{\ka-1}{\ka}\right|  \\
+\frac{(\ka-r)^4}{\ka} \left( \frac{\ka+4r}{r^2}\right)  \ln\left|\frac{\ka-r}{\ka}\right|{\bigg]}  + 
{\bigg[}\ka\to-\ka{\bigg]} \bigg\}.
\end{multline}
where $\kappa=K/\kf$ and {the notation $\kappa\to -\kappa$ means that $\kappa$ has to be replaced by its opposite
in {\sl all} terms within the main square brackets.}
This function is plotted in Fig.~\ref{fig:re-eff} for different values of the impurity-to-fermion mass ratio $r$.
The locus of points where the moduli in the argument of the $\ln$ functions vanish forms the boundaries of the same physical regions 
discussed for $\Im f(K/\kf)$, and sketched in Fig.~\ref{fig:regions}. We have numerically observed that Eq.~(\ref{eq:Real_f}) is a monotonically 
decreasing function of $\ka$.
It has a vanishing derivative at the origin, $\Re f'(0)=0$, where it reaches its maximum value 
\be
\label{eq:Real_f_0}
\Re f(0) = \frac{9r(1-r^2+2r^2\ln r)}{4(r^2-1)^2}.
\ee
Eq.~(\ref{eq:Real_f_0}) was obtained in \cite{Walecka,BishopNucl} in the 
context of a $\Lambda$-particle in a Fermi sea of nucleons with $\nu=4$ spin-isospin states, hence
an extra factor $\nu$ in those references, {and also in \cite{Albus} for a Bose-Fermi mixture.}
Eq.~(\ref{eq:Real_f}) vanishes at large $K/\kf$:
\be
\label{eq:re-large}
\Re f(\ka) \underset{\ka \gg 1+r}{=} \frac{r}{4\ka^2} + O(1/\ka)^4.
\ee
This indicates that, for large $K/\kf$ and up to second-order in $g$,  the dispersive effect of the Fermi sea on the energy of the impurity reduces
to the mean-field term $\rho g$.
Remarkably, Eq.~(\ref{eq:Real_f}) has a finite limit for a unit mass ratio
\begin{multline}
\label{eq:eff_r1}
\lim_{r \to1} \Re f(\ka) = \frac{3}{2}-\frac{3}{40}\bigg\{{\bigg[}\frac{11}{2} + \ka^2 + (2\ka^3+4\ka^2+6\ka+3)\\
\times\frac{(\ka-1)^2}{\ka}\ln|\ka-1| -2\ka^4\ln|\ka|{\bigg]} + {\bigg[}\ka\to-\ka{\bigg]}\bigg\},
\end{multline}
and as well as for $\Im f(K/\kf)$ the crossover region is absent in this special case $M=m$. In the weakly attractive limit, $g\to0^-$, the second-order effect of 
the Fermi sea on the energy of the impurity moving through it can then be summarized as follows for an arbitrary mass ratio: On the real (imaginary) part of the resulting complex energy the effect of the medium is maximal (minimal) when the impurity is at rest, $\KK=\OO$, while the effect of the medium is minimal (maximal) when the impurity is moving fast.

A related observable of interest is the effective mass of the polaron. Since the impurity is moving with momentum $\hbar \KK$,
one can define the effective mass $m_\parallel^*(\KK)$ along the direction of $\KK$, and the effective mass $m_\perp^*(\KK)$ along the direction
perpendicular to $\KK$~\cite{Mermin}. 
$m_\parallel^*(\KK)$ is related to the second derivative of the complex energy 
Eq.~(\ref{eq:ener_pol}) with respect to the impurity wave number $K =\ka \kf$ as follows:
\be
\label{eq:invmeff}
\frac{M}{m_\parallel^*(\KK)} \underset{g\to0^-}{=} 1 + \frac{r}{2} \left(\frac{\rho g}{\Ef}\right)^2 f''(\ka) +O(g^3),
\ee
and $m_\perp^*(\KK)$ is related to the first derivative:
\be
\label{eq:invmeff_perp}
\frac{M}{m_\perp^*(\KK)} \underset{g\to0^-}{=} 1 + \frac{r}{2} \left(\frac{\rho g}{\Ef}\right)^2 \frac{f'(\ka)}{\ka} +O(g^3).
\ee
When $K\to0$, Eqs.~(\ref{eq:invmeff}, \ref{eq:invmeff_perp}) have the same limit, which coincides with the
result of~\cite{Bishop}, and one recovers the usual
rotational symmetry of the effective mass tensor.
At non-zero $K$, Eq.~(\ref{eq:invmeff_perp}) is a differentiable function of $K$, whereas Eq.~(\ref{eq:invmeff}) presents interesting singularities
and is non-differentiable in $\ka=1$ as shown in Figs.~\ref{fig:im-eff}, \ref{fig:re-eff}.
Furthermore, for $r=1$, $\Re f''(\ka)$ presents a logarithmic divergence $\propto \ln|\ka -1|$ in $\ka=1$, as can be  
seen in Eq.~(\ref{eq:eff_r1}). 
This suggests that  for $r=1$, $M/m_\parallel^*(\KK)$ cannot be Taylor expanded in powers of $g$ at the point $\ka=1$,
as e.g.\  the function $g^2\ln|g|$, and a non perturbative approach must be used \cite{TCTC}.
Instead, $\Im f''(\ka)$ has a jump $J = \Im f''(1^+)-\Im f''(1^-) = 9\pi/4$ in $\ka=1$ for $r=1$.

\section{Physical interpretation}
We now interpret the boundaries between regions  in Fig.~\ref{fig:regions}. Intuitively, singularities 
in the polaron complex energy may originate from the fact that (i) the energy denominator in Eq.(\ref{eq:ener_boite}) can vanish, and (ii) the domains of
variation of $\kk$ and $\qq$ in Eq.(\ref{eq:ener_boite}) have sharp boundaries $k=q=\kf$. 
One thus investigates the form of the Fourier space domain 
where the energy denominator vanishes for $\kk$ and $\qq$ {\sl both} at the Fermi surface. For $K<\kf$, it is found that
this domain supports all values of $\theta\in [0,\pi/2]$, where $\theta$ is the angle between $\KK$ and $\kk-\qq$. 
For $K>\kf$, on the contrary, the allowed values of $\theta$ range from $\arccos(\kf/K)$ to $\pi/2$. This designates
$K=\kf$ as a first peculiar line in the plane $M/m$ -- $K/\kf$. Note that $\pi-2\theta$ is the scattering angle
of an incoming impurity of wave vector $\KK$ on the Fermi sea when {\sl both} the resulting scattered fermion and hole are at the Fermi surface.
To obtain the second peculiar line in Fig.~\ref{fig:regions},  we introduce the additional idea that the energy denominator {\sl as well as}  its first order derivatives
with respect to $\kk$ and to $\qq$ vanish at $k=q=\kf$, which indeed leads to  $K=r\kf$.

\section{Narrow Feshbach resonance}
In a recent experiment~\cite{Zaccanti} impurities of $^{40}$K were immersed in a Fermi sea of $^6$Li, where the $s$-wave interaction
between the impurity and each fermion has a non-negligible range due to the presence of a narrow Feshbach resonance.
Such a situation differs from the one we have considered so far: A new length appears in the system, namely
the Feshbach length $R_*$, which at the experimental density~\cite{Zaccanti} was of the order of $R_*\simeq \kf^{-1}$
and thus not negligible.
This physical situation can be described by a two-channel model~(see e.g. \cite{TCPRA} and references therein).
For simplicity we keep in the two-channel model the same lattice model as for the single-channel Hamiltonian (\ref{eq:H0},\ref{eq:V}). 
The Hamiltonian of the system then becomes $\hat{\mathcal{H}} = \hat{\mathcal{H}}_0+\hat{W}+E_\mathrm{mol}\hat{N}_\mathrm{mol}$, with
\begin{eqnarray}
\label{eq:H0_2C}
\hat{\mathcal{H}}_0 &=& \hat{H}_0+\sum_\kk \frac{\varepsilon_\kk}{1+r}\bb^\dag_\kk\bb_\kk, \\
\hat{W}     &=& \frac{\Lambda}{\sqrt{\mathcal{V}}}\sum_{\kk_1,\kk_2, \kk_3} \delta^{\rm mod}_{\kk_1+\kk_2,\kk_3}
\left( \bb^\dag_{\kk_3} \uu_{\kk_2} \dd_{\kk_1}+\mathrm{h.c.}\right) 
\label{eq:V_2C}
\end{eqnarray}
and $\hat{N}_\mathrm{mol} = \sum_\kk  \bb^\dag_\kk\bb_\kk$ is the number of closed-channel molecules operator.
Here $\bb_\kk$ annihilates a closed-channel molecule with wave vector $\kk$, kinetic energy $\hbar^2 k^2/[2(M+m)]=\varepsilon_\kk/(1+r)$ and internal energy $E_\mathrm{mol}$.
The relation between $E_\mathrm{mol}$, the interchannel coupling $\Lambda$, the Feshbach length $R_*$ and the $s$-wave scattering length $a$ is \begin{equation}
E_\mathrm{mol}=-\frac{\Lambda^2}{g_0} \;\; \mathrm{and} \;\; R_*=\frac{\pi\hbar^4}{\Lambda^2\mu^2},
\end{equation}
where $g_0$ is still given by Eq.~(\ref{eq:bare}). For a fixed lattice spacing $b$ and Feshbach length $R_*$, one sees that in the weakly attractive limit 
$a\to0^{-}$ the energy of the closed-channel molecule diverges, $E_\mathrm{mol}\to+\infty$. Therefore, in such a limit we introduce the hermitian projector 
$\hat{P}$ onto the low-energy subspace in which closed-channel molecules are absent, and the projector $\hat{Q}=\hat{1}-\hat{P}$. 
Within this low-energy subspace one can derive an exact effective Hamiltonian depending on a complex energy $z$~\cite{CCT}:
\begin{equation}
\label{eq:Heff}
\hat{\mathcal{H}}_\mathrm{eff}(z) = \hat{P} \hat{\mathcal{H}} \hat{P} +  \hat{P} \hat{W} \hat{Q} \frac{\hat{Q}}{z\hat{Q}-\hat{Q}\hat{\mathcal{H}}\hat{Q}} \hat{Q}\hat{W} \hat{P}.
\end{equation}
The complex energy $E_{\rm FS}(N)+\Delta E$ of the impurity must then be (in the thermodynamic limit and under appropriate analytic continuation) an eigenvalue of $\hat{\mathcal{H}}_\mathrm{eff}(E_{\rm FS}(N)+\Delta E)$, which constitutes an implicit equation. Here we solve it perturbatively
up to order $g_0^2$ by making an asymptotic expansion of~(\ref{eq:Heff}) in powers of $E_\mathrm{mol}$ to obtain
in the subspace of momentum $\hbar \KK$:
\begin{multline}
\label{eq:Heff02}
\hat{\mathcal{H}}_\mathrm{eff}(E_{\rm FS}(N)+\Delta E) \underset{g\to0^-}{=} \hat{P} \hat{H} \hat{P} \\
-\frac{1}{E_\mathrm{mol}^2} \hat{P}\hat{W} \hat{Q}[E_\KK+E_{\rm FS}(N)-\hat{\mathcal{H}}_0]
\hat{Q} \hat{W} \hat{P} + O(g_0^3),
\end{multline}
where we used $-E_\mathrm{mol}^{-1} \hat{P} \hat{W} \hat{Q} \hat{W} \hat{P} = \hat{P} \hat{V} \hat{P}$ and the fact that
$\Delta E=E_\KK+O(g_0)$.
Remarkably, one recovers as the first term the single-channel model Hamiltonian (\ref{eq:H0},\ref{eq:V}) modified by the second term of (\ref{eq:Heff02}),
which is an effective-range correction proportional to $R_*$, that vanishes, as it should, when $R_*\to0$. 
It remains to apply the usual second-order perturbation theory in $g_0$ to the effective Hamiltonian~(\ref{eq:Heff02}).
This amounts to adding to the already calculated single channel result the contribution of the second term of (\ref{eq:Heff02})
simply treated to first order in perturbation theory, which is real since this term is hermitian.
After thermodynamic and zero-lattice spacing limits we obtain:
\be
\label{eq:tild_ener_pol}
\Delta E(\KK) \underset{g\to0^-}{=} E_\KK + \rho g + \frac{(\rho g)^2}{\Ef} \tilde{f}(K/\kf) + O(g^3),
\ee
where
\be
\label{eq:ft}
\tilde{f}(\ka) = f(\ka) - \frac{3\pi}{2}\frac{r}{(1+r)^3} \left(\ka^2+\frac{3r^2}{5}\right) \kf R_*.
\ee
The correction to $f(\ka)$ is a smooth function of $\ka$ so that the regions of Fig.~\ref{fig:regions}
remain unchanged for $\tilde{f}$.
 
At this point a comparison with the hard-sphere result \cite{Bishop} at $K=0$ is possible with some care.
First, one formally identifies the effective range $r_e=-2 R_*$ of the two-channel model with the one
$2a/3$ of the hard-sphere model, which turns the second term of (\ref{eq:ft}) into an energy correction
of order $a^3$. Second, this energy correction must appear in the theory of \cite{Bishop} as
a single $T$-matrix vertex, that is in the contribution $\epsilon_1$, see Eq.~(3.5) of \cite{Bishop}.
Third, one must include the fact that the hard-sphere interaction also scatters in the $p$-wave channel,
with a scattering volume $\mathcal{V}_s=a^3/3$ for the convention of \cite{JonaLasinio}, which
leads to an energy shift given by  first order perturbation theory applied to the pseudo-potential of \cite{Jolicoeur}.
Then we find consistency between our result and the one of \cite{Bishop} at $K=0$.

\section{Validity conditions} For our perturbative treatment to apply, the impurity-to-fermion scattering
must be in the Born regime, with a scattering amplitude $f_\krel=-a/[1+a(i\krel +\krel^2R_*)] \simeq -a$ at the relative wave number 
$\krel = \mu |\KK/M-\qq/m| \le (K+r\kf)/(1+r)$, so that
\be
\krel |a| \ll 1 \;\;\;\; \mbox{and} \;\;\;\; \krel^2 |a| R_* \ll 1. 
\ee
However, this is not sufficient when $\tilde{K}\equiv K/(1+r) \gg\kf$: In this regime one can use the $T$-matrix formalism \cite{Bishop} to first order, neglecting 
the initial momentum of the fermions, to obtain $\Delta E(\KK)-E_\KK \simeq - \rho (2\pi \hbar^2/\mu) f_{\tilde{K}}$. Requiring that, in the real part of this result, the terms of order three in $a$ are small as compared to the $a^2$ ones
gives
\be
a^2 \tilde{K}^2 |1-\tilde{K}^2 R_*^2| \ll \left| \frac{\kf a}{6\pi(1+r)} \frac{\kf^2}{\tilde{K}^2} -a \tilde{K}^2 R_*\right|,
\ee
the first term in the right-hand side coming from Eq.~(\ref{eq:re-large}).

\section{Experimental feasibility}
To determine the experimental relevance of 
this work we 
refer to the experimental conditions~\cite{Zaccanti}, where for $\kf |a| < 2$ the real part 
(imaginary part) of the polaronic complex energy can be measured with a precision of $5\times10^{-3}\Ef$ ($10^{-4}\Ef$). 
Then our theory for a moving polaron may be tested experimentally  for a finite scattering length when two conditions hold
on $\kf |a|$: (i) the second-order correction in Eq.~(\ref{eq:tild_ener_pol}) must be sufficiently large to be observed, and (ii) the third-order correction $\Delta E^{(3)}$, which is 
neglected in Eq.~(\ref{eq:tild_ener_pol}), must be sufficiently small. 
Extending the perturbative method used to derive Eq.~(\ref{eq:Heff02}) up to order $g_0^3$ we obtain
\begin{multline}
\label{eq:correction03}
\Delta E^{(3)}(\KK) = g^3 \int'\frac{d^3q}{(2\pi)^3} 
\left\{ \frac{1}{\Lambda^2}\left(E_\KK+\varepsilon_\qq -\frac{\varepsilon_{\KK+\qq}}{1+r}\right) \right.\\
\left. - \int \frac{d^3k}{(2\pi)^3}\left[\frac{2\mu}{\hbar^2k^2}-\frac{Y(k-\kf)}{F_{\kk,\qq}(\KK)}\right] \right\}^2 -g^3\frac{\rho^2}{\Lambda^2}\\
-g^3\int \frac{d^3k}{(2\pi)^3}\left[\int'\frac{d^3q}{(2\pi)^3} \frac{Y(k-\kf)}{F_{\kk,\qq}(\KK)}\right]^2,
\end{multline}
with $F_{\kk,\qq}$ defined in Eq.~(\ref{eq:Fkq_def}). The first (third) contribution originates from the subspace with one hole and two
particles (one particle and two holes), and the second contribution originates from a conspiracy between the mean-field and 
the effective-range terms within the subspace with one hole. At $\KK=\OO$ and $R_*=O(a)$, Eq.~(\ref{eq:correction03}) 
reproduces~\cite{Bishop}.

\begin{figure}[t]
\begin{center}
\includegraphics[width=1.0\linewidth,clip=]{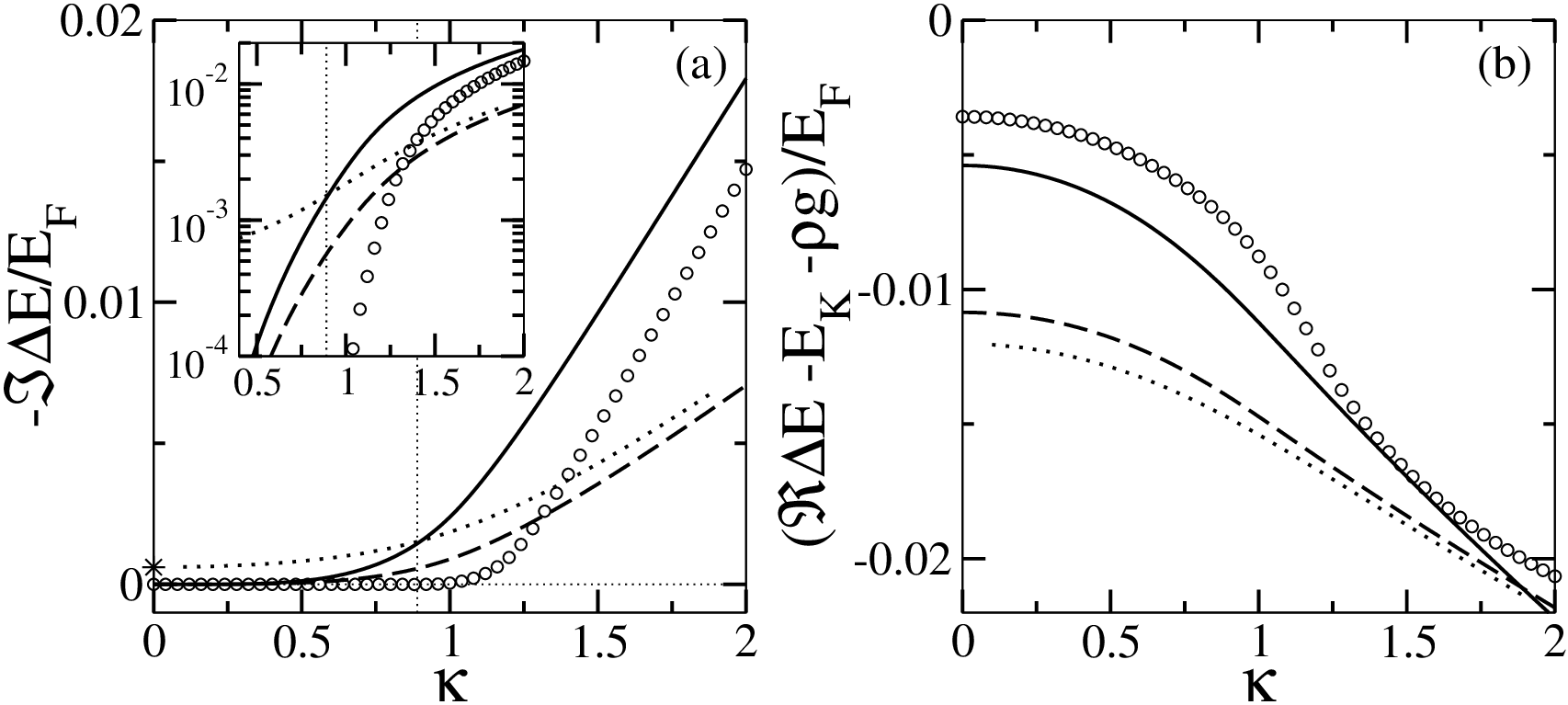}
\caption{For a $^{40}$K impurity interacting with a Fermi sea of $^6$Li ($r=6.6439$) on a narrow Feshbach resonance with $\kf R_* = 1$
as in \cite{Zaccanti}, and $\kf a=-0.46$. Imaginary part (a), and real part (b) of the complex energy 
counted with respect to the {mean energy of the ideal Fermi gas}, as a function of its reduced wave number $\ka=K/\kf$. In (b) the mean-field prediction 
$E_\KK+\rho g$ was subtracted for clarity. Dashed (solid) lines: Perturbative calculation up to order two (three) in $g$.
Circles: The single particle-hole ansatz. On the left of the vertical dotted line of 
(a) the ansatz incorrectly predicts a zero imaginary part. Inset in (a): magnification with vertical axis in log-scale.
{Perturbative calculation up to $g^2$ generalized at finite temperature $T$, for $k_B T=0.05 \Ef$ as in \cite{Ketterle}: 
dotted line; star: Eq.~(15) of \cite{Albus}.}
}
\label{fig:experimental01}
\end{center}
\end{figure}

We have performed explicit calculations for $\kf a=-0.46$, which is intermediate between weakly and strongly interacting
regime: In Fig.~\ref{fig:experimental01}b we show that, for the real part of the complex energy, the condition of experimental observability 
is indeed satisfied over the useful range $0\leq K \leq 2\kf$,
with a reasonably small correction due to $\Re \Delta E^{(3)}(\KK)$; we further test the real part of our perturbative result 
up to $n=3$ (solid line) against the real part resulting from the single particle-hole ansatz (12) of \cite{TCPRA} (circles), showing reasonable agreement.
Instead for the imaginary part in Fig.~\ref{fig:experimental01}a, while $\Im \Delta E^{(2)}(\KK)$ is measurable within the experimental precision~\cite{Zaccanti}
for $K > 0.5 \kf$, the correction due to $\Im \Delta E^{(3)}(\KK)$ is not negligible, and the full perturbative result up to $n=3$ differs qualitatively
from the ansatz (12) of \cite{TCPRA}. The reason for this difference has to be sought in the ansatz, which
wrongly predicts a zero decay rate of a moving impurity with wave number $K<K_c$, due to the fact that the effect of the interaction
is incorrectly described in the particle-hole continuum, which gives a wrong (here zero) value to the lower border of that continuum 
(see Eq.~(24) of \cite{TCPRA}). To find $K_c$, one simply note that 
within the ansatz approach a real solution to the moving polaron is then always possible if $\Delta E(\KK)<0$~\cite{TCPRA}, which at the mean-field level
leads for $g\to 0^-$ to 
$K_c \sim \kf [2(r+1)\kf |a|/(3 \pi)]^{1/2}$,
already a good approximation for Fig.~\ref{fig:experimental01}.

{We have also included in Fig.~\ref{fig:experimental01} a finite-temperature calculation, by generalizing 
Eq.~(\ref{eq:integral},\ref{eq:Heff02}) for fermions at thermal equilibrium with density $\rho$, and by numerically 
evaluating the resulting integrals. At $k_B T=0.05\Ef$ as in \cite{Ketterle}, the correction to the real part 
is small, but the imaginary part no longer tends to zero at $K=0$ \cite{ZLan}.}

\section{Conclusion}
In the weakly attractive limit, both for broad and narrow Feshbach resonances, we have calculated analytically, up to second order in the interaction strength $g$, the energy, the decay rate, and the effective masses of an impurity moving 
with momentum $\hbar\KK$ through a fully polarized Fermi sea, extending known results for $K=0$ to finite momenta. At that order, our results show the existence of four regions separated by singularities
of system's observables, as e.g.\ the effective mass. We have characterized the degree and the physical origin of these singularities, which
lays in the impurity-to-fermion scattering at the Fermi surface.
For realistic values of the parameters we have shown that our second order correction is experimentally observable with the radio-frequency spectroscopic
technique. We have also given an integral expression for the third-order complex energy, so as to estimate the neglected terms in the second-order theory.

\bigskip
We thank M. Zaccanti and M. Cetina for discussions. We acknowledge fundings from the ERC FERLODIM N.228177 and the Marie Curie INTERPOL N.298449.

\end{document}